\documentclass[useAMS,usenatbib]{mn2e}

\usepackage{graphicx}
\usepackage{ifpdf}
\usepackage{epstopdf}
\usepackage{url}
\usepackage{multirow}

\title[CCSN Rate Synthesis Within 11 Mpc]{Core-Collapse Supernova Rate Synthesis Within 11 Mpc}
\author[Lin Xiao and J.J. Eldridge]{Lin Xiao$^{1}$\thanks{E-mail: lin.xiao@auckland.ac.nz} and J.J. Eldridge$^{2}$\thanks{E-mail: j.eldridge@auckland.ac.nz}\\
$^{1}$,$^{2}$Department of Physics, University of Auckland, NZ}
\begin{document}

\pagerange{\pageref{firstpage}--\pageref{lastpage}} \pubyear{2015}
\maketitle

\label{firstpage}

\begin{abstract}
The 11 Mpc $ {\rm H\alpha} $ and Ultraviolet Galaxy (11HUGS) Survey traces the star formation activity of nearby galaxies. In addition within this volume the detection completeness of core-collapse supernovae (CCSNe) is high therefore by comparing these observed stellar births and deaths we can make a sensitive test of our understanding of how stars live and die. In this paper, we use the results of the Binary Population and Spectral Synthesis (BPASS) code to simulate the 11HUGS galaxies' $ {\rm H\alpha} $ and far-ultraviolet (FUV) star formation rate indicators (SFRIs) and simultaneously match the core-collapse supernova (CCSN) rate. We find that stellar population including interacting binary stars makes little difference to the total CCSN rate but increases the $ {\rm H\alpha} $ and FUV fluxes for a constant number of stars being formed. In addition they significantly increase the predicted rate of type Ibc supernovae (SNe) relative to type II SNe to the level observed in the 11HUGS galaxies. We also find that instead of assuming a constant star formation history (SFH) for the galaxies our best fitting models have a star formation rate (SFR) that peaked more than 3 Myrs ago.

\end{abstract}


\section{Introduction}

Core-collapse supernovae (CCSNe) are the explosive events that mark the death of massive stars ($ \ga 8{\rm M_{\odot}} $) \citep{2009MNRAS.395.1409S}. They are classified into hydrogen-rich events, Type IIP/L/n/b, and hydrogen-deficient events, type Ib/c, according to the presence or absence of hydrogen lines in their spectrum \citep{1997ARA&A..35..309F}. This is directly related to the composition of the outer layers of a progenitor star at the moment of explosion. Type II supernova (SN) progenitors have kept their hydrogen rich outer envelope intact, while type Ib/c have lost the hydrogen envelope. The most direct evidence linking the star that explodes to the type of SN is from direct detection of the progenitor in pre-explosion images \citep[e.g.][]{2007ApJ...661.1013L,2009ARA&A..47...63S}. Non-detections of progenitors in pre-explosion images for type Ibc SN were used by \citet{2013MNRAS.436..774E} to indicate the important effect of interacting binary stars on determining the CCSN progenitor diversities and local CCSN rate. This idea was later reinforced by the probable detection of a binary system progenitor for the type Ibc SN iPTF13bvn \citep{2013ApJ...775L...7C,2014AJ....148...68B,2015MNRAS.446.2689E}.  \\
\\However, it is difficult to directly detect CCSN progenitors because the event must occur in a galaxy within about 30 Mpc that has been imaged by the Hubble Space Telescope before explosion. Therefore indirect methods must be exploited to gain further insight into CCSNe. The most useful methods include the relative rates of different CCSN types \citep[e.g.][]{1992ApJ...391..246P,1998A&A...333..557D,2008Ap.....51...69H,2008MNRAS.384.1109E,2009A&A...503..137B,
2011MNRAS.412.1522S}, the stellar populations surrounding the site of CCSNe \citep[e.g.][]{2014ApJ...791..105W} and the location of CCSNe within their host galaxies \citep[e.g.][]{ 2006Natur.441..463F,2006A&A...453...57J,2008ApJ...687.1201K,2012MNRAS.424.1372A}.
\\
\\One other interesting constraint on CCSNe and their progenitors can be obtained by exploiting the link between stellar deaths and stellar births. This was attempted by \citet{2012A&A...537A.132B}. They used the 11HUGS galaxy catalogue that lists the $ {\rm H\alpha} $ and FUV fluxes for a large number of galaxies within 11 Mpc. These fluxes are commonly used tracers of recent star formation \citep{1998ARA&A..36..189K}. The amount of $ {\rm H\alpha} $ and FUV emitted are determined by the most massive hot stars that evolve quickly and explode as CCSNe a few million years after their formation. \citet{2012A&A...537A.132B} were able to show that the observed $ {\rm H\alpha} $ and FUV fluxes from the 11HUGS galaxies were consistent with the observed CCSN rate for the galaxies if the minimum stellar mass for a CCSN to occur was $ 8{\rm M_{\odot}} $, in agreement with results from CCSN progenitor detection in \citet{2009MNRAS.395.1409S}.\\
\\This study was repeated by \citet{2013ApJ...769..113H} who tried a slightly more sophisticated modelling of the 11HUGS galaxies. Rather than using the canonical expected SFR conversion factors from \citet{1998ARA&A..36..189K} they estimated the expected $ {\rm H\alpha} $, FUV flux and CCSN rate from the latest stellar models including rotation calculated by \citet{2012A&A...537A.146E}. Over the last few decades the understanding of how rotation affects stellar evolution has significantly advanced \citep[see reviews by][]{2000A&A...361..159M, 2012ARA&A..50..107L}. The general effect is to increase the main-sequence life time and temperature of massive stars as they evolve so lead to more $ {\rm H\alpha} $ and FUV flux being emitted for the same SFR. \citet{2013ApJ...769..113H} found that it is possible that there could be a CCSN rate excess as well as that the observations disfavoured the rotating models as they made this SN rate excess worse.\\
\\There is a second important factor left for our stellar population models that is the effect of interacting binaries. When two stars are orbiting each other in a binary-star system the opportunity for mass transfer between the two stars arises. The more massive star typically evolves faster than the less massive star, fills its Roche-Lobe and transfers mass to the companion star. This produces more hot and luminous stars and therefore has similar observed consequence as stellar rotation does \citep[e.g.][]{1998A&ARv...9...63V,2008MNRAS.384.1109E,2012AcASn..53..274Z}. However the situation may be more complex as there is evidence that the most rapid rotation only occurs in binary systems \citep{2013ApJ...764..166D}. \\
\\In this paper we continue the work of \citet{2012A&A...537A.132B} and \citet{2013ApJ...769..113H} by attempting to match the observed $ {\rm H\alpha} $, FUV fluxes and CCSN rate in the 11HUGS galaxies. We will use the Binary Population and Spectral Synthesis (BPASS, \url{http://bpass.auckland.ac.nz}) code that accounts for the impact of interacting binaries on stellar populations (Eldridge \& Stanway 2009, 2012). The BPASS code has been tested against a number of different observables that are dependant on the evolution of massive stars. Our aim is to apply a simple test to BPASS models to reproducing the CCSN rate, $ {\rm H\alpha} $ and FUV fluxes. Our work concentrates on the effect of interacting binary stars on the CCSN rate of their host galaxies. We also include two other refinements to our models. First we reproduce the chemical compositions of the 11HUGS galaxies based on the B-band magnitude of nearby galaxies sample rather than using a Solar composition population. Second we allow the SFH to vary rather than assuming only a constant SFR.\\
\\The structure of this paper is arranged as follows. In Section 2, we introduce the 11HUGS and CCSN samples and how we determine the metallicity of the 11HUGS galaxies. In the following Section 3, we summarise the effect of interacting binary stars on CCSN rate and SFRIs we find from the BPASS code's results. We also outline how we use these BPASS results to construct our synthetic galaxy models. In Section 4, we compare our models to observations before finally discussing our results and presenting our conclusions in Section 5.

\section{Galaxy and CCSNe samples}
\begin{table}
\caption{Nearby galaxy sample from the 11 HUGS survey \citep{{2004AAS...205.6004L},2012A&A...537A.132B} and CCSNe sample from \citet{2013MNRAS.436..774E}: Numbers of galaxies, the total $ {\rm L_{H\alpha}/10^{43}erg \, s^{-1} }$, $ {\rm L_{UV}/10^{28}erg \, s^{-1}Hz^{-1} }$ and CCSN rate ($ {\rm R_{CC}/yr^{-1} }$). $ {}^{1} $The ${\rm L_{UV}} $ for Sample A is an estimation with respect to the relative ratio of $ {\rm H\alpha} $ and UV flux from Sample B and C. }
\begin{center}
                      \begin{tabular}{l|c|c|c}
                        \hline\hline
                        Parameter & Sample A & Sample B & Sample C\\
                        \hline
                        $ {\rm N_{gal}} $  & 383 & 312 & 167\\
                        
                        \hline
                        $ {\rm L_{H\alpha}/10^{43}erg \, s^{-1} }$ &  $ {\rm 1.1 \pm 0.05} $  & $ {\rm 0.9 \pm 0.05} $ &  $ {\rm 0.7 \pm 0.04} $ \\
                        $ {\rm L_{UV}/10^{28}erg \, s^{-1}Hz^{-1} }$ & $ {}^{1} {106 \pm 8}$ &$ {\rm 88 \pm 6} $ & $ {\rm 67 \pm 4} $\\
                        \hline
                        $ {\rm N_{CC}} $  & 17 & 15 & 14\\
                        $ {\rm N_{SN I}} $  & 4 & 4 & 3\\
                        $ {\rm N_{SN II}} $  & 13 & 11 & 11\\
                        $ {\rm R_{CC}/yr^{-1} }$ & $ {\rm 1.13 \pm 0.27 } $ & $ {\rm 1 \pm 0.26 } $ & $ {\rm 0.93 \pm 0.25 } $\\
                        \hline
                        
\end{tabular}
\end{center}
\label{tab:local galaxies}
\end{table}

The galaxy samples we adopt in this paper are drawn from  \cite{2012A&A...537A.132B}. There are three different samples: 383 galaxies with measured flux in $ {\rm H\alpha} $ (sample A), 312 galaxies with measured fluxes in $ {\rm H\alpha} $ and FUV (sample B) and 167 galaxies with measured fluxes in $ {\rm H\alpha} $, FUV and Total Infrared (TIR) (sample C). We list the observed fluxes of these samples in Table \ref{tab:local galaxies}. We broadly assume that the relative ratio of $ {\rm H\alpha} $ and FUV flux is constant. Therefore applying the averaged flux ratio of $ {\rm H\alpha} $ and FUV of Sample B and C to Sample A, we estimate the FUV luminosity of Sample A: $ {\rm L_{UV} = (106 \pm 8) \times 10^{28}erg \, s^{-1}Hz^{-1} }$. \\
\\ In addition, we list the details of the 17 CCSNe discovered within 11 Mpc from 1998 to 2012 in Table \ref{tab:CCSNe_11Mpc}. It is directly based on the work of \citet{2012A&A...537A.132B}, where they picked out all identified SNe occurred in galaxy sample A of 11HUGS survey from 1885 to 2010 from the Asiago SNe catalogue \citep{2008yCat....102024B}. We limit our SNe sample to only core-collapse events. We also extend this CCSN list up to 2012 using the SN list from \citet{2013MNRAS.436..774E}. This list of SNe was compiled by searching the catalogue maintained by the International Astronomical Union (IAU) Central Bureau for Astronomical Telegrams (CBET)\footnote{http://www.cbat.eps.harvard.edu/lists/Supernovae.html}. It was searched for all SNe discovered in the fifteen-year period from 1 Jan 1998 to 30 March 2012. The host galaxies of these selected SNe have a recessional velocity (corrected for Local Group infall on Virgo) of 2000 $ \rm{km s^{-1}} $ or less, corresponding to a distance limit of 28 Mpc for $ \rm{H_{0} = 72 km s^{-1} Mpc^{-1}} $. \\
\\All CCSNe used in our work is in the field of 11HUGS survey and related to the adopted SFRIs. Of these there are 13 type II SNe and 4 type Ibc SNe. Among all the CCSNe host galaxies, the host galaxy of SN 2005at is in Sample A and B but not in Sample C, and the host galaxy of both SN 2005af and SN 2011ja is only in Sample A. It is important to consider whether our sample of SN is complete and whether there might have been some CCSNe unobserved. For example some CCSNe might occur in very dusty regions of galaxies and remain unobserved or some SNe might form black holes and remain unobserved \citep[e.g.][]{2015arXiv150402635S}. However modern SN searches are unlikely to have missed many SNe. In addition if any SNe are unobservable due to dust, their associated $ {\rm H\alpha} $ and FUV emission will also be decrease thus in this work should not contribute to the observed SFRIs. Although in light of these concerns we can suggest that our CCSN rate is at least a robust lower limit.
\begin{table}
\caption{CCSNe discovered within 11 Mpc from 1998 to 2012 as listed by \citet{2013MNRAS.436..774E}. ${}^{1}$SN 2005at host galaxy NGC 6744 is in Sample A and B not in Sample C ,$ {}^{2} $SN 2005af and SN 2011ja host galaxy NGC 4945 is only in Sample A.}

\begin{center}
                      \begin{tabular}{l|c|c|c|c}
                        \hline
                        \hline
                        SN  & Gal. & Type & $ {\rm M_{B}} $ & D/Mpc \\ 
                        \hline
                       2002ap & NGC 628 & Ic & -19.65 & 7.3\\ 
                       2002bu & NGC 4242 & IIn & -18 & 7.4 \\ 
                       2002hh& NGC 6946 & IIP & -20.65& 5.9 \\ 
                       2003gd & NGC 628 & IIP & -19.65 & 7.3 \\ 
                       2004am & NGC 3034 & IIP & -19.08 & 3.53 \\ 
                       2004dj & NGC 2403 & IIP & -18.77 & 3.22 \\ 
                       2004et & NGC 6946 & IIP & -20.65 & 5.9 \\ 
                       2005cs & NGC 5194 & IIP & -20.7 & 8 \\ 
                       2007gr & NGC 1058 & Ic & -18.2 & 9.2 \\ 
                       2008ax & NGC 4490 & IIb & -19.39 & 8 \\
                       2008bk & NGC 7793 & IIP & -18.41 & 3.91 \\ 
                       2009hd & NGC 3627 & IIP & -19.61 & 10.05 \\ 
                       2011dh & NGC 5194 & IIb & -20.7 & 8 \\ 
                       2012fh & NGC 3344 & Ic & -18.8 & 6.6 \\ 
                        \hline
                       $ {}^{1} $ 2005at& NGC 6744 & Ic & -20.9 & 9.4 \\ 
                       \hline
                       $ {}^{2} $ 2005af& NGC 4945 & IIP & -20.6 & 4.08 \\ 
                       $ {}^{2} $ 2011ja& NGC 4945 & IIP & -20.6 & 4.08 \\ 
                       \hline
\end{tabular}
\end{center}
\label{tab:CCSNe_11Mpc}
\end{table}

\section{Method}
Our method to reproduce the 11HUGS $ {\rm H\alpha} $ flux, UV flux and CCSN rate falls into three steps. First a metallicity classification of the 11HUGS galaxies is applied so that we can blend our models of different metallicities and obtain a more realistic simulation of 11HUGS sample. Second we describe the BPASS stellar population model that we use to construct our models, and discuss the effect of interacting binary stars on CCSN rate and SFRIs of star-forming galaxies. Finally we outline our method to create a synthetic galaxy model with various star-formation histories.
\subsection{Metallicity Classification}
\begin{table}
\caption{Metallicity classification of Sample B galaxies: the fraction of total galaxies, ${\rm H\alpha }$ flux, FUV flux and number of CCSNe when the 11HUGS galaxies are divided into the 5 metallicities of BPASS.}
\begin{center}
                      \begin{tabular}{l|c|c|c|c|c|c}
                        \hline
                        \hline
                        & Z  & ${\rm [O/H]} $ & $ {\rm f_{gal} }$& ${\rm f_{H\alpha} }$& ${ \rm f_{FUV} }$& $ {\rm f_{CCSN} }$ \\
                        \hline
                       \multirow{5}{*}{$ {\rm T04} $}&0.001 & $ 7.5 $ & 0.178 & 0.003 & 0.002 & 0\\
                        & 0.004 & $ 8.1  $ & 0.451 & 0.038 & 0.033 & 0\\
                        & 0.008 & $ 8.4 $ & 0.197 & 0.134 & 0.096 & 0.057\\
                        & 0.020 & $ 8.8  $ & 0.117 & 0.324 & 0.253 & 0.531 \\
                        & 0.040 & $ 9.0  $ & 0.057 & 0.501 & 0.616 & 0.412\\
                        \hline
                         \multirow{5}{*}{$ {\rm KK04} $}&0.001 & $ 7.5 $ & 0.083 & 0.001 & 0.001& 0\\
                        & 0.004 & $ 8.1  $ & 0.292 & 0.009 & 0.009 & 0\\
                        & 0.008 & $ 8.4 $ & 0.317 & 0.052 & 0.04 & 0\\
                        & 0.020 & $ 8.8  $ & 0.225 & 0.339 & 0.258 & 0.353 \\
                        & 0.040 & $ 9.0  $ & 0.083 & 0.599 & 0.692 & 0.647\\
                        \hline
                         \multirow{5}{*}{$ {\rm PP04} $}&0.001 & $ 7.5 $ & 0 & 0 & 0 & 0\\
                        & 0.004 & $ 8.1  $ & 0.498 & 0.052 & 0.043 & 0\\
                        & 0.008 & $ 8.4 $ & 0.419 & 0.427 & 0.324 & 0.294\\
                        $ {\rm O3N2 } $& 0.020 & $ 8.8  $ & 0.079 & 0.521 & 0.633 & 0.706\\
                        & 0.040 & $ 9.0  $ & 0.003 & 0 & 0 & 0\\
                        \hline
                        
\end{tabular}
\end{center}
\label{tab:metallicity groups}
\end{table}
Previous studies have only considered reproducing the stellar populations in 11HUGS samples by using only Solar metallicity composition for the stellar populations. However it is well known that synthetic population results depend on metallicity even when interacting binaries are included. Therefore we take account of the varying chemical compositions of the 11HUGS galaxies. We estimate the galaxies' metallicity by using the luminosity-metallicity (L-Z) relation of the SDSS galaxies provided by \cite{2004ApJ...613..898T}. This shows generally good agreement with other local samples of galaxies. The relation is,
\begin{eqnarray}
   [O/H] & = & 12 + \log(O/H) \nonumber \\
         & = & -(0.185\pm 0.001)M_{B} + (5.238\pm 0.018)
\end{eqnarray}
The ratio of oxygen and hydrogen [O/H], is a common measurement of galaxy metallicity and able to be estimated by the B-band magnitude of the host galaxy taken from \citet{2011ApJS..192....6L}. However, this L-Z relation is strongly affected by dust attenuation and the choice of which strong-line abundance calibrations are used as discussed in \cite{2008ApJ...681.1183K}. To quantify the influence of dust, \cite{2004ApJ...613..898T} correct the galaxy luminosities for intrinsic attenuation using the attenuation curve of \cite{2000ApJ...539..718C} and assume that the stars experience one third of the attenuation measured in nebular gas. To account for the effect of using different strong-line abundance calibrations, we use the work of \cite{2008ApJ...681.1183K} who derived conversion relations between different strong-line calibrations. Here we use their conservation relations between Tremonti model (T04) and two other models of \cite{2004ApJ...617..240K} model (KK04) and \cite{2004MNRAS.348L..59P} model (PP04 O3N2). These two models are the possible extremes to the T04 model. Therefore by comparing to these we can estimate how our results depend on the metallicity classification assumed.\\
\\BPASS has five different metallicities for its model populations, ranging from one twentieth to twice Solar. To determine how much star formation occurs at different metallicity environments we compare the galaxies' metallicity to the input values for the BPASS models and calculate the fraction of total observed galaxy population ($ {\rm f_{gal} }$), ${\rm H\alpha} $ flux (${\rm f_{H\alpha} }$), FUV flux (${ \rm f_{FUV} }$) and CCSN rate ($ {\rm f_{CCSN} }$) at each metallicity. For this classification we use Sample B because it is large enough to cover more than 80 per cent of the total sample and has both $ {\rm H\alpha} $ and FUV observations available. Our metallicity classifications with respected to different strong-line calibrations are presented in Table \ref{tab:metallicity groups}.\\
\begin{table*}
\caption{Predicted flux of $ {\rm H\alpha} $ and FUV at SFR of $ {\rm {1M_{\odot}}yr^{-1}} $, $ {\rm {10M_{\odot}}yr^{-1}} $, $ {\rm {50M_{\odot}}yr^{-1}} $, $ {\rm {100M_{\odot}}yr^{-1}} $ with estimated uncertainty, due to stochastic effect of constructing the synthetic galaxy models from individual star clusters whose masses were selected at random.} 
\begin{center}
                      \begin{tabular}{l|c|c|c|c|c|c|c|c|c}
                        \hline
                        \hline
                       SFR& Z  & \multicolumn{2}{c}{$ {\rm L_{Ha}/(10^{41}erg \, s^{-1})} $ }& \multicolumn{2}{c}{$ {\rm L_{FUV}/(10^{28}erg \, s^{-1}Hz^{-1})} $} & \multicolumn{2}{c}{$ {\rm R_{CCSN}/(10^{-2}yr)} $} & \multicolumn{2}{c}{$ {\rm R_{Ibc}}/R_{II} $}\\
                        \hline
                      & & single & binary & single & binary & single & binary & single & binary\\
                      \hline
                        \multirow{4}{*}{1$ {\rm {M_{\odot}}/yr} $}& 0.001 & $ 3.1 \pm 1.2 $ & $ 4.4 \pm 1.5 $& $ 1.3 \pm 0.2 $  & $ 1.8 \pm 0.2 $ &  $ 1.2 \pm 0.2  $ & $ 1.1 \pm 0.2  $ & $ 0.003 \pm 0.003 $ & $ 0.46 \pm 0.04  $\\
                       & 0.004 & $ 2.5 \pm 0.9 $ & $ 5.7 \pm 1.3 $ &$ 1.1 \pm 0.2 $  & $ 1.9 \pm 0.2 $ &$ 1.2 \pm 0.2 $ & $ 1.2 \pm 0.2  $ & $ 0.07 \pm 0.04 $ & $ 0.62 \pm 0.09  $\\
                       & 0.008& $ 2.2 \pm 0.9 $ & $ 3.7 \pm 1.3 $ &$ 1.1 \pm 0.2 $ & $ 1.6 \pm 0.3 $ & $ 1.2 \pm 0.2 $ & $ 1.1 \pm 0.2  $ & $ 0.11 \pm 0.06  $& $ 0.61 \pm 0.11  $ \\
                       & 0.020 & $ 1.6 \pm 0.7 $ & $ 2.7 \pm 1.0 $ &$ 1.0 \pm 0.2 $& $ 1.4 \pm 0.3 $ & $ 1.0 \pm 0.2 $ & $ 1.0 \pm 0.2  $ & $ 0.2 \pm 0.09  $ & $ 0.64 \pm 0.13  $\\
                       & 0.040 & $ 1.2 \pm 0.5 $ & $ 2.0 \pm 0.7 $ &$ 0.9 \pm 0.2 $  & $ 1.3 \pm 0.3 $ & $ 0.8 \pm 0.2 $ & $ 0.8 \pm 0.2  $ & $ 0.23 \pm 0.11  $ & $ 0.74 \pm 0.17  $\\
                        \hline
                        \multirow{4}{*}{10$ {\rm {M_{\odot}}/yr} $}& 0.001 & $ 31.2 \pm 4.0 $ & $ 44 \pm 5.0 $ & $ 12.7 \pm 0.7 $ & $ 17.9 \pm 0.8 $ & $ 12.0 \pm 0.6 $ & $ 11.0 \pm 0.5 $ & $ 0.003 \pm 0.0008 $ & $ 0.46 \pm 0.01 $ \\
                       & 0.004 & $ 25.5 \pm 3.4 $ & $ 57.8 \pm 4.7 $ & $ 11.3 \pm 0.6 $ & $ 19.2 \pm 0.9 $ & $ 11.9 \pm 0.6 $ & $ 11.8 \pm 0.6 $ & $ 0.06 \pm 0.01 $ & $ 0.63 \pm 0.03 $\\
                       & 0.008& $ 22.0 \pm 2.9 $ & $ 37.0 \pm 4.0 $ & $ 10.8 \pm 0.7 $ & $ 16.4 \pm 0.9 $ & $ 12.0 \pm 0.6 $ & $ 11.2 \pm 0.5 $ & $ 0.11 \pm 0.02 $ & $ 0.61 \pm 0.04 $ \\
                       & 0.020 & $ 16.1 \pm 2.4 $ & $ 26.9 \pm 3.3 $ & $ 9.7 \pm 0.7 $ & $ 14.0 \pm 0.9 $ & $ 9.8 \pm 0.5 $ & $ 9.9 \pm 0.5 $ & $ 0.19 \pm 0.03 $ & $ 0.64 \pm 0.04 $ \\
                       & 0.040 & $ 11.7 \pm 1.7 $ & $ 19.9 \pm 2.3 $ & $ 8.6 \pm 0.6 $ & $ 12.9 \pm 0.8 $ & $ 8.2 \pm 0.5 $ & $ 8.4 \pm 0.5 $ & $ 0.22 \pm 0.04 $ & $ 0.73 \pm 0.05 $ \\
                        \hline
                        \multirow{4}{*}{50$ {\rm {M_{\odot}}/yr} $}& 0.001 & $ 155 \pm 9 $ & $ 219 \pm 11 $ & $ 63 \pm 1.3 $ & $ 89 \pm 1.6 $ & $ 60 \pm 1.1 $ & $ 53 \pm 1.1 $ & $ 0.003 \pm 0.0004 $ & $ 0.46 \pm 0.006 $ \\
                       & 0.004 & $ 128 \pm 7 $ & $ 290 \pm 10 $ & $ 57 \pm 1.3 $ & $ 96 \pm 1.7 $ & $ 60 \pm 1.1 $ & $ 59 \pm 1.2 $ & $ 0.06 \pm 0.006 $ & $ 0.63 \pm 0.01 $ \\
                       & 0.008& $ 110 \pm 6 $ & $ 185 \pm 8 $ & $ 54 \pm 1.2 $ & $ 82 \pm 1.6 $ & $ 60 \pm 1.2 $ & $ 56 \pm 1.1 $ & $ 0.11 \pm 0.008 $ & $ 0.61 \pm 0.01 $ \\
                       & 0.020 & $ 81 \pm 5 $ & $ 135 \pm 7 $ & $ 49 \pm 1.5 $ & $ 70 \pm 1.9 $ & $ 49 \pm 1.1 $ & $ 49 \pm 1.0 $ & $ 0.19 \pm 0.01 $ & $ 0.64 \pm 0.02 $ \\
                       & 0.040 & $ 59 \pm 4 $ & $ 100 \pm 5 $ & $ 43 \pm 1.3 $ & $ 65 \pm 1.7 $ & $ 41 \pm 1.3 $ & $ 42 \pm 1.3 $ & $ 0.22 \pm 0.02 $ & $ 0.73 \pm 0.03 $ \\
                        \hline
                        \multirow{4}{*}{100$ {\rm {M_{\odot}}/yr} $}& 0.001 & $ 310 \pm 11 $ & $ 438 \pm 14 $ & $ 127 \pm 1.5 $ & $ 179 \pm 1.8 $ & $ 119 \pm 1.4 $ & $ 106 \pm 1.3 $ & $ 0.003 \pm 0.0003 $& $ 0.46 \pm 0.004 $ \\
                       & 0.004 & $ 252 \pm 9 $ & $ 574 \pm 11 $ & $ 113 \pm 1.5 $ & $ 192 \pm 1.9 $ & $ 120 \pm 1.5 $ & $ 118 \pm 1.5 $ & $ 0.06 \pm 0.002 $ & $ 0.62 \pm 0.008 $ \\
                       & 0.008& $ 219 \pm 7 $ & $ 369 \pm 10 $ & $ 107 \pm 1.7 $ & $ 164 \pm 2.2 $ & $ 119 \pm 1.2 $ & $ 112 \pm 1.0 $ & $ 0.10 \pm 0.005 $ & $ 0.60 \pm 0.01 $ \\
                       & 0.020 & $ 162 \pm 6 $ & $ 271 \pm 8 $ & $ 98 \pm 1.6 $ & $ 140 \pm 2.2 $ & $ 98 \pm 1.7 $ & $ 99 \pm 1.6 $ & $ 0.19 \pm 0.009 $ & $ 0.64 \pm 0.01 $\\
                       & 0.040 & $ 119 \pm 5 $ & $ 201 \pm 7 $ & $ 86 \pm 1.6 $ & $ 130 \pm 2.1 $ & $ 82 \pm 1.6 $& $ 83 \pm 1.3 $ & $ 0.22 \pm 0.01 $ & $ 0.73 \pm 0.02 $ \\
                        \hline
\end{tabular}
\end{center}
\label{tab:HaNUV}
\end{table*}

\noindent We take T04 model as our standard metallicity classification and find lower metallicity galaxies dominate in number as they are mainly dwarf galaxies. They contribute little
to the FUV flux and a small but significant fraction of the $ {\rm H\alpha} $ flux. None of the CCSNe occur in the lowest metallicity galaxies. The galaxies from LMC (Z=0.008) to twice Solar
metallicity (Z=0.04) however contribute most of the light and CCSNe in our sample. We use the relative fractions of the FUV flux at different metallicities to create our model population because of a better agreement between FUV flux and the CCSN rate. The $ {\rm H\alpha} $ flux peaks towards lower metallicities and is sensitive to a shorter period of star formation, only the last 10 Myrs rather than the longer timescale for CCSN and FUV flux of 100 Myrs. Therefore the FUV flux better represent the averaged SFR during the period where SN progenitors will be born. We would expect the low metallicity dwarf galaxies to be younger with more $ {\rm H\alpha} $ dominated star burst than high metallicity galaxies. Also recent work by \citet{2015arXiv150201347B} suggested that the amount of $ {\rm H\alpha} $ emission that escapes depend on the intensity of star formation in a galaxy. From their result we expect more $ {\rm H\alpha} $ emission to escape from dwarf galaxies but less $ {\rm H\alpha} $ emission to escape from giant galaxies which have more diffuse star-forming regions. However the main reason why the $ {\rm H\alpha} $ and FUV fluxes are greater is that lower metallicity stars are hotter and therefore emit more $ {\rm H\alpha} $ flux relative to FUV flux.  \\ 
\\The variation of emitted fluxes with metallicity can be seen in our model populations shown in Table \ref{tab:HaNUV}. It presents the predicted $ {\rm H\alpha} $ and FUV flux at different metallicities from single-star and binary-star populations with star formations of 1$ {\rm {M_{\odot}}/yr} $, $ {\rm {10M_{\odot}}yr^{-1}} $, $ {\rm {50M_{\odot}}yr^{-1}} $, $ {\rm {100M_{\odot}}yr^{-1}} $ along with uncertainties in the prediction. These fluxes are derived through our synthetic galaxy model which is discussed in detail in Section 3.3. The SFR is contributed from total galaxy's star formation which vary with stochastic effect from constructing galaxy by individual star clusters. Therefore at given SFR the uncertainties of flux, CCSN rate and CCSN relative ratio, are highly dependent on how many star clusters are required. In addition, the uncertainties decrease with increasing SFR as stochastic effect decrease. The FUV fluxes for both single and binary models, decline slightly as metallicity increases compared to the $ {\rm H\alpha} $ flux which changes by a factor of 2. Moreover, FUV flux experience similar declining as CCSN rate does when metallicity increase. As for the relative ratio of type Ibc to type II SN, it gradually increases with the increasing metallicity but keeps almost constant at given metallicity whatever the value of total SFR is. In addition, compared to single-star model binary-star model produced more $ {\rm H\alpha} $ and FUV flux and increased the relative ratio of type Ibc and type II SN, but made little difference in total CCSN rate. This will be discussed in detail in section 3.2. On balance, FUV flux is our best choice for this metallicity classification for the 11HUGS galaxies.\\
\begin{table*}
\caption{$ {\rm H\alpha} $ and FUV flux, CCSN rate and ratio predictions at 1$ {\rm {M_{\odot}}/yr}$ of star formation rate following different metallicity relations.}
\begin{center}
                      \begin{tabular}{l|c|c|c|c|c|c|c|c}
                        \hline
                        \hline
                        Metallicity & \multicolumn{2}{c}{$ {\rm L_{Ha}/(10^{41}erg \, s^{-1})} $ }& \multicolumn{2}{c}{$ {\rm L_{FUV}/(10^{28}erg \, s^{-1}Hz^{-1})} $} & \multicolumn{2}{c}{$ {\rm R_{CCSN}/(10^{-2}yr)} $} & \multicolumn{2}{c}{$ {\rm R_{Ibc}}/R_{II} $}\\
                        \cline{2-9}
                         Model& single & binary & single & binary & single & binary & single & binary \\
                        \hline
                        T04& $ 1.39 \pm 0.59 $ & $ 2.41 \pm 0.77 $ & $ 0.95 \pm 0.17 $ & $ 1.45 \pm 0.30 $& $ 0.97 \pm 0.22 $ & $ 0.98 \pm 0.23 $ & $ 0.14 \pm 0.12 $ & $ 0.60 \pm 0.15 $ \\
                        KK04 & $ 1.34 \pm 0.57 $ & $ 2.29 \pm 0.79 $ & $ 0.96 \pm 0.17 $ & $ 1.45 \pm 0.28 $ & $ 0.98 \pm 0.21 $ & $ 0.99 \pm 0.22 $ & $ 0.15 \pm 0.10 $ & $ 0.59 \pm 0.15 $\\
                        PP04 O3N2 & $ 1.82 \pm 0.77 $ & $ 3.26 \pm 1.06 $ & $ 1.07 \pm 0.16 $& $ 1.62 \pm 0.30 $ & $ 1.08 \pm 0.19 $ & $ 1.12 \pm 0.18 $ & $ 0.14 \pm 0.08 $ & $ 0.61 \pm 0.12 $\\
                        \hline

\end{tabular}
\end{center}
\label{tab:metallicity prediction}
\end{table*}
\\The choice of different metallicity calibrations can lead to large variations in the classification of 11HUGS sample as shown in Table \ref{tab:metallicity groups}. Compared to the T04 model, the KK04 model is biased towards higher metallicities in all aspects: observed galaxy fraction, the fraction of $ {\rm H\alpha} $ and FUV flux and observed CCSN fraction. Conversely the PP04 model shows a preference to lower metallicities. Even though there are still around 70 per cent of galaxies in KK04 model laying below solar metallicity, over 90 percent of $ {\rm H\alpha} $ and FUV flux and total CCSN are from solar and twice-solar metallicities. On the other hand PP04 O3N2 model push more galaxies below solar metallicity and almost half of the total $ {\rm H\alpha} $ flux, 40 per cent of FUV flux and 30 per cent of CCSN are given from galaxies below solar metallicity. These differences in metallicity classifications are finally resulting in the variations of our metallicity-mixed models shown in Table \ref{tab:metallicity prediction}.\\
\\We find, compared to T04, the KK04 model favours higher metallicities, decreasing the $ {\rm H\alpha} $ flux slightly and no effects on FUV flux for both single-star and binary-star models. While the PP04 model that is biased to low metallicities, enhances the $ {\rm H\alpha} $ emission by a third and slightly increases FUV emission. This result is consistent with what we get from a single composition as shown in Table \ref{tab:HaNUV}, with more $ {\rm H\alpha} $ and FUV fluxes from lower metallicities and larger variations in $ {\rm H\alpha} $ flux than FUV flux. However different metallicity classifications only lead to a slight variation in CCSN rate and the relative ratios from different metallicity models even keep almost constant. This is because for all the metallicity classifications the CCSN rate is dominated by either solar (PP04 
O3N2 model) or twice-solar (KK04 model) metallicity galaxies. Table \ref{tab:HaNUV} shows a little difference in the CCSN rates and the relative ratios between these two single metallicity models. In summary, we can match the observations better by changing the metallicity compositions of galaxies. In our following work we adopt T04 model as a standard to make metallicity classification of 11HUGS sample. These supplementary models we have created allow us to estimate how uncertain our predictions are due to assumed metallicities of the 11HUGS galaxies.

\begin{figure*}
\centering

\includegraphics[width=8.8cm]{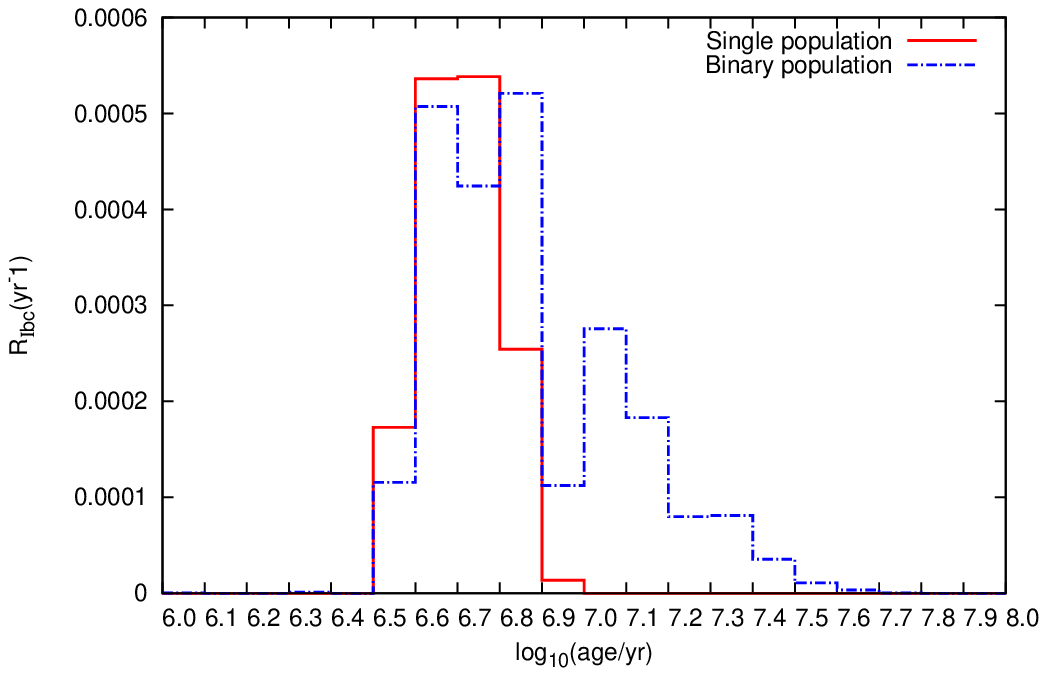}
\includegraphics[width=8.8cm]{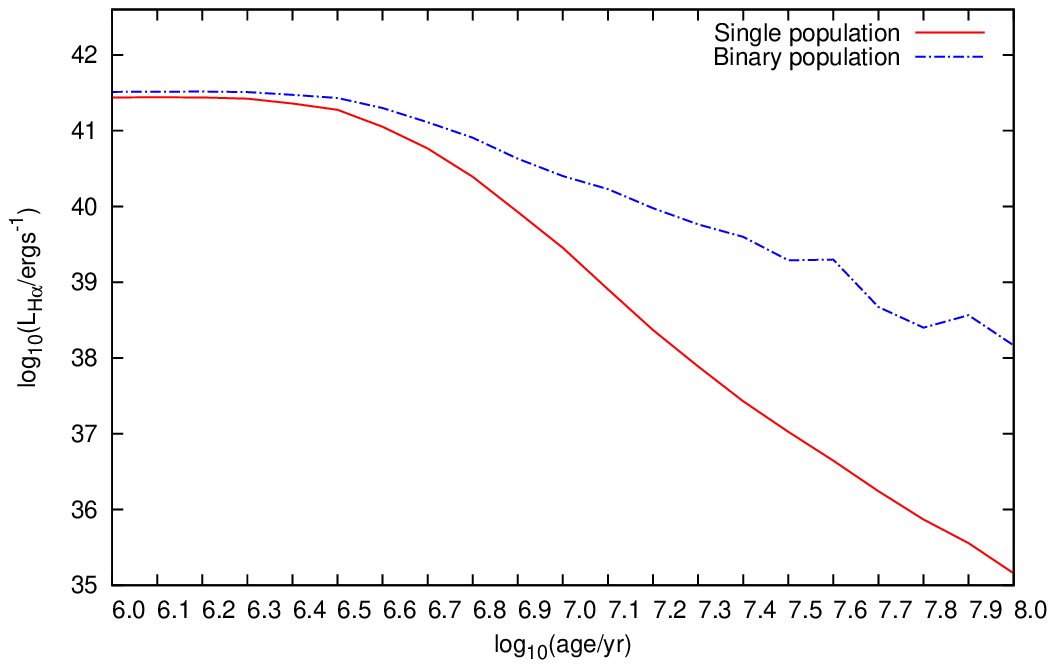}
\includegraphics[width=8.8cm]{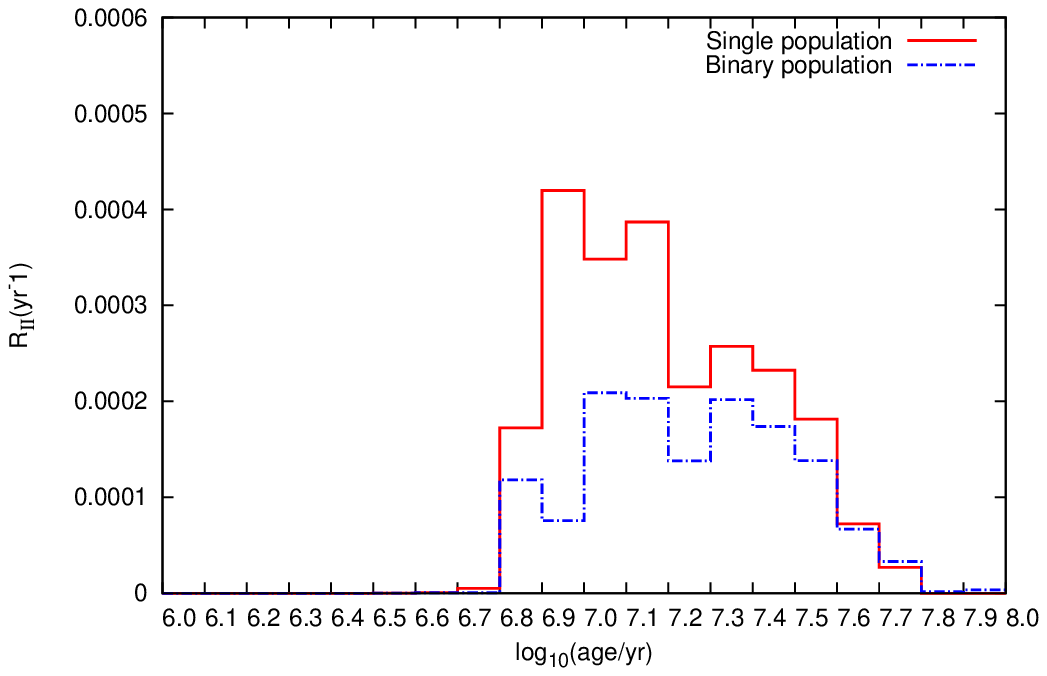}
\includegraphics[width=8.8cm]{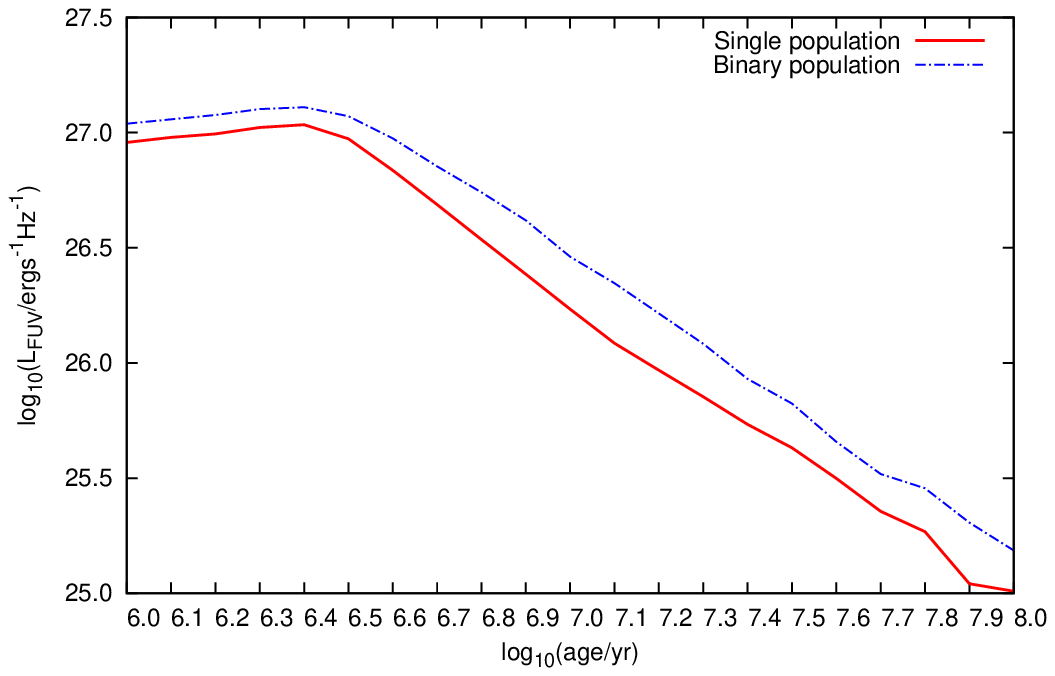}
\centering
\caption{Distributions of CCSN rate in terms of stellar age for type Ibc SNe in the left top panel and for type II SNe in the left bottom panel. The relations of $ {\rm H\alpha} $ and FUV flux across stellar age are shown in right top panel and right bottom panel separately. The red solid lines are from single-star model and blue dashed lines are from binary-star model. } \label{fig:FluxSNe}
\end{figure*}

\subsection{Binary Effect on CCSN Rate and SFRIs}

We construct our galaxy model using the results of BPASS, a population-synthesis model based on a detailed stellar evolution code calculated in \citet{2008MNRAS.384.1109E}. There are two sets of detailed evolution model, single stars and an extensive set of binary stars, to produce a realistic synthetic stellar population as observations show approximately 70 per cent of massive stars have their evolution affected by binary interactions \citep[e.g.][]{2012Sci...337..444S}. These evolution models are combined with an atmosphere model to predict emitted spectrum of the stars. 
Moreover, a composite stellar spectrum is calculated for a range of ages. This process is described in detail in Eldridge \& Stanway (2009, 2012).\\
\\To calculate the CCSN rate, first we determine whether a star produces CCSN and which types CCSN occur at the end of its evolution. We use the same requirements for both single-star population and binary-star population. A star with final mass greater than 2${\rm M_{\odot}} $ and carbon-oxygen core mass over 1.38${\rm M_{\odot}} $ are taken to experience a CCSN. All the identified CCSN progenitors are then divided into type II SNe with hydrogen in the final model and type Ibc SNe without hydrogen. Therefore, the fate of a star at given initial mass is determined by its mass loss during its lifetime. Compared to single-star model, mass transfer in an interacting binary systems can remove the outer hydrogen envelope from a star's surface and also lead to mass gain from the companion star. This result is a difference in the CCSN rate of different types as shown in Figure \ref{fig:FluxSNe}. The predicted rates are from the composite model obtained by mixing the five metallicity populations. For both single-star and binary-star model, type II SNe arise from similar stellar population in terms of stellar age arranging from 6 Myrs to 60 Myrs, because more massive stars that die earlier have strong stellar winds that will remove the hydrogen for both sets of models. The binary model however have a reduced occurrence of type II SNe and increased the probability of type Ibc SNe. In addition, a type Ibc SN progenitor can only occur after a lifetime of 10 Myrs from an interacting binary star. When it comes to the total CCSN rate, binary populations make little difference, but as we have demonstrated it plays an important role in changing the relative ratio of type Ibc to type II SNe. \\
\\Interacting binaries also have a great effect on both $ {\rm H\alpha} $ and FUV fluxes.  Although the effect diverges with age. The $ {\rm H\alpha} $ line from a star-forming galaxy is produced by hydrogen gas that have been photonionized by hot massive stars. This emission is contributed by stars with stellar ages less than 16 Myrs \citep{1998ARA&A..36..189K,2009ApJ...691..115G}. Therefore, we derived $ {\rm H\alpha} $ luminosity by taking account of all the Lyman continuum photons $ {\rm N(LyC)} $ below the wavelength of ${\rm \lambda = 912\mathring{A} } $ and follow the hydrogen recombination relation provided by \citet{2006agna.book.....O}:
\begin{equation}
L(H\alpha) = 1.36 \times 10^{-12} N({\rm LyC}) \, { \rm  erg \, s^{-1} }
\end{equation}
\noindent While the FUV flux arises from the continuum spectrum at the wavelength between ${\rm 1556\mathring{A} } $ to ${\rm 1576\mathring{A} } $ directly from the stars themselves spectrum. Stars contribute to this with ages up to approximately 100 Myrs \citep{1998ARA&A..36..189K,2009ApJ...691..115G}.  \\
\\How the $ {\rm H\alpha} $ and FUV fluxes vary with stellar age of our composite metallicity population are shown in the right two panels of Figure \ref{fig:FluxSNe}. We see that the binary-star population produces more flux than single-star population. These are in agreement with the greater values predicted by \citet{2012A&A...541A..49Z}. The extra emission occurs because of the binary interactions creating more hot stars with their hydrogen envelope stripped and mass-gain in binary systems. The FUV fluxes follow each other with the binary populations being only 0.1 to 0.2 dex greater at any age. \\
\\ A greater difference occurs for the $ {\rm H\alpha} $ flux. After 3 Myrs which is the life duration of a typical H II region \citep{2013MNRAS.428.1927C}, the $ {\rm H\alpha} $ flux from binary-star model does not decay as rapidly as single-star model. This again is due to more longer-lived stars but also more hot massive stars at later times to continue to contribute to the $ {\rm H\alpha} $ flux. In addition, the short lifetime of the typical H II region implies that binary-star models have only a small effect on observed H II regions. However over longer timescale the amount of diffuse $ {\rm H\alpha} $ emission from a galaxy might represent less star-formation than previously thought.

\subsection{Our Synthetic 11HUGS Model}
\begin{figure*}
\centering
\includegraphics[width=250pt]{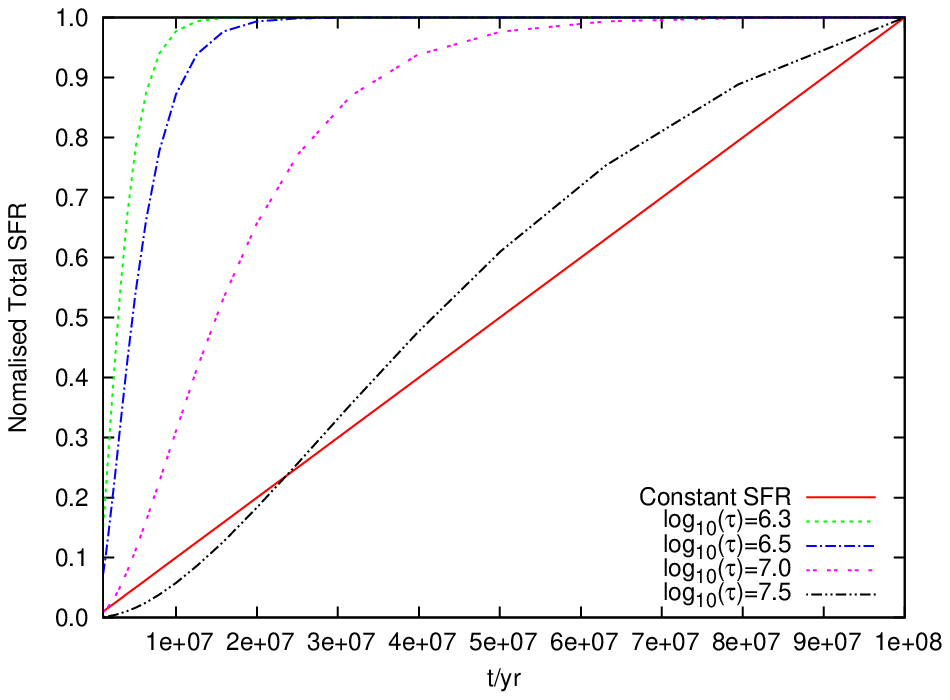}
\includegraphics[width=250pt]{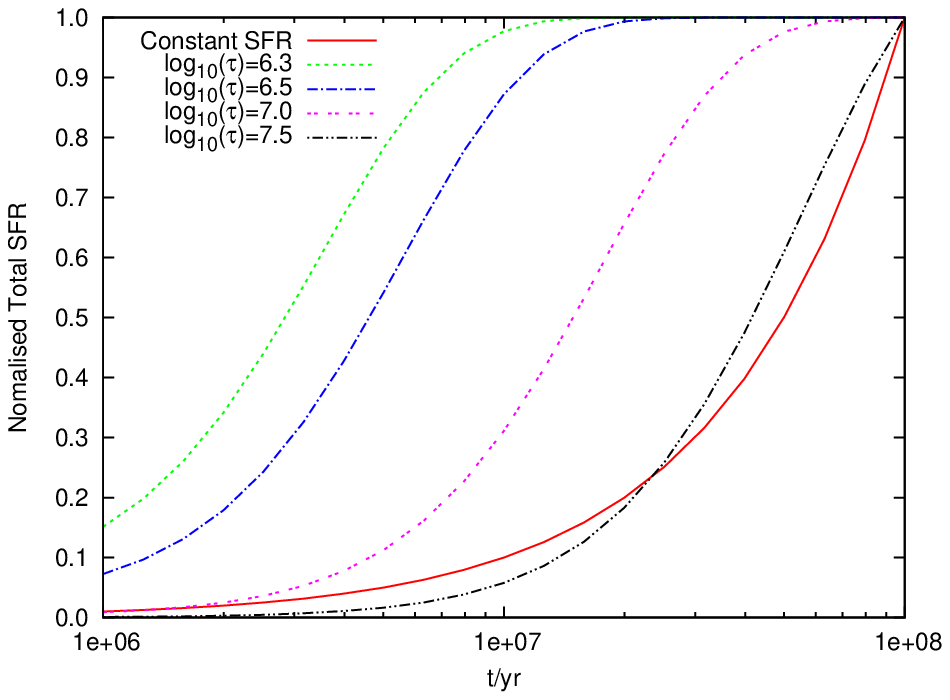}
\centering
\caption{Delayed SFR as a function of time under 4 different time scale parameter with logarithm value of 6.3, 6.5, 7.0 ,7.5 respectively, and compare with constant SFH in red solid lines. The right panel is in linear time scale and the left panel is in logarithm time scale.} \label{fig:DecayedSFR}
\end{figure*}
Using the BPASS results described above, we construct our synthetic 11HUGS galaxies model as follows. We follow the method of \citet{2012MNRAS.422..794E} by creating galaxies from individual star clusters from a range of masses between 50$ {\rm M_{\odot}} $ and $ {\rm 10^{6}M_{\odot}} $. We pick the masses at random according to a cluster mass function from \citet{2003AJ....126.1836H} of the form:
\begin{equation}
\frac{dN}{dM} \propto M^{-2} 
\end{equation}
\noindent Rather than uniformly distributing the age of clusters we set their age randomly too. For example to reproduce a constant star formation we allow the cluster to have any age from 0 to 10 Gyrs ago. With each cluster's age and mass we then select and scale the appropriate BPASS output and add it into our total predicted spectrum and current CCSN rate of model galaxies. After repeated work as above for different metallicity galaxy populations we mix the model galaxies based on the metallicity classification of 11HUGS sample and obtain a composite 11 HUGS model. With this method of creation we have a prediction of SFRIs appearance and CCSN rate from stellar populations spanning a SFR from $ {\rm 10^{-3}M_{\odot}yr^{-1}} $ to $ {\rm 100M_{\odot}yr^{-1}} $. At low SFRs our results do predict a range of possible appearances due to stochastic effects which can be seen in Table \ref{tab:HaNUV} \citep[e.g.][]{{2009MNRAS.395..394P},{2011ApJ...741L..26F},{2012MNRAS.422..794E},{2012ApJ...744...44W}}, however in the amount of SFR of the 11HUGS sample are at a level beyond stochastic effects.\\
\\A further variation to our 11HUGS model is that we have also created models with a delayed SFH as in \cite{2010ApJ...725.1644L}. The SFR, $ \Psi(t,\tau) $, in the delayed SFH is given by 
\begin{equation}
\Psi(t, \tau) \propto \frac{t}{\tau^{2}}e^{-t/ \tau}
\end{equation}
where t is the time since the onset of star formation and $ {\rm \tau} $ is the star formation time scale parameter. In our work we allow $ {\rm \tau} $ to vary from 1 Myrs to 30 Myrs. Compared to constant SFH, delayed SFH has its dominated star-burst time range around $ {\rm \tau} $ where we can expect relatively high SFR appears. This time range is directly dependent on the choice of $ {\rm \tau} $. Increasing $ {\rm \tau} $ increases the proportion of older stars relative to the younger stars, and therefore alters the predicted results of our synthetic models. As shown in Figure \ref{fig:DecayedSFR}, SFH with lower values of $ {\rm \tau} $ experience a sharper increase at earlier times. As $ {\rm \tau} $ increases, delayed SFH approach to constant SFH evolving similarly. From the plot in logarithm time scale, we can have a more clearly look at its dominated star-burst period. Here, the position with the largest increasing slope correspond to a star-burst peak which is around $ {\rm \tau} $. After the peak SFH decline gradually and a lower value of $ {\rm \tau} $ speed up this decline. As s result, increasing the value of $ {\rm \tau} $ does not only delay the star-burst peak time, but also extend the duration of the star formation. Moreover, we can associate our predicted CCSN rate to a stellar population with its dominated star-burst period known.\\
\\Our synthetic galaxy models under a delayed SFH and constant SFH predicting CCSN rate and full synthetic spectrum can be compared directly to the observed 11HUGS galaxies.   

\section{Comparison With 11HUGS}

\subsection{Constant SFH Model}
\begin{figure*}
\centering
\includegraphics[width=250pt]{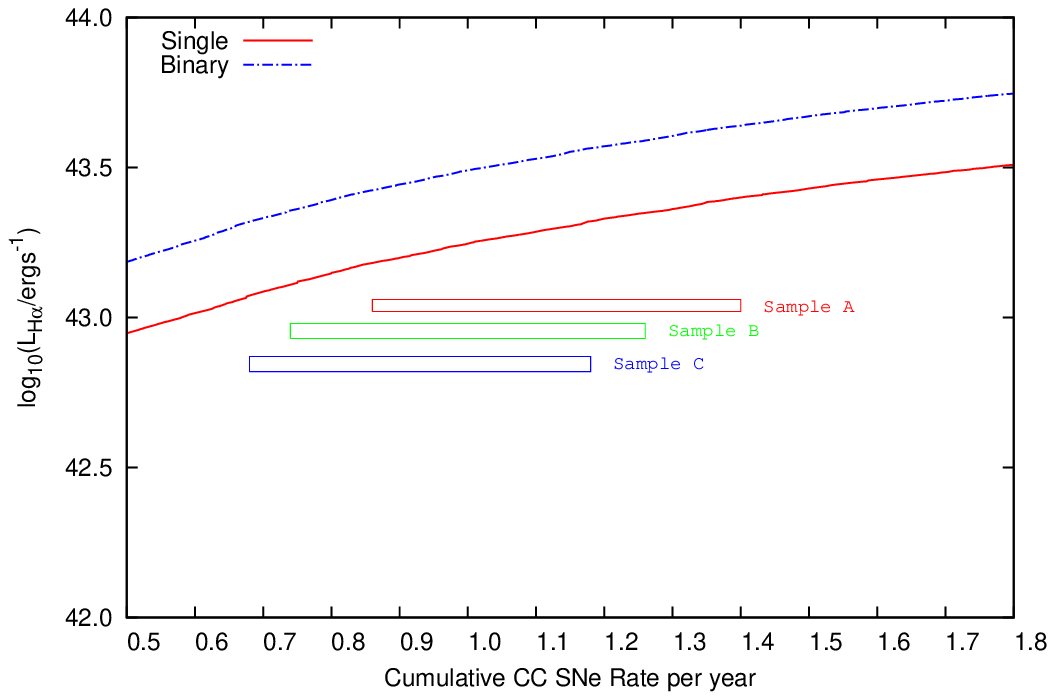}
\includegraphics[width=250pt]{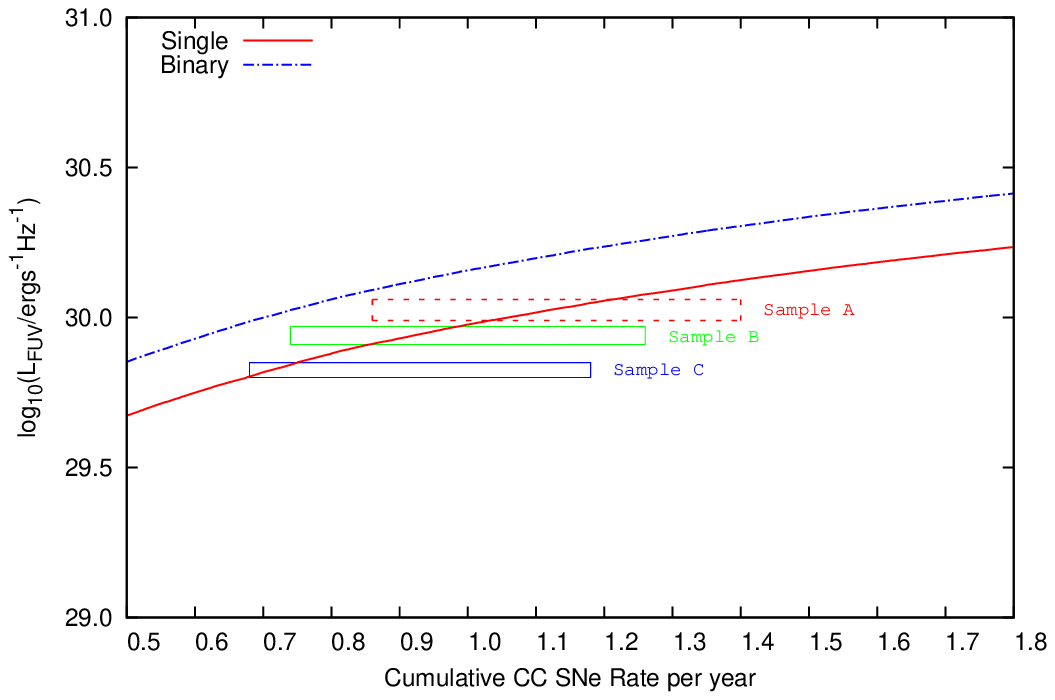}
\centering
\caption{Constant star foramtion model: $ {\rm H\alpha} $ and FUV flux as a function of cummulative CCSN rate. The red solid lines are from single-star model and blue dashed lines are from binary-star model. The colored boxes show the observed flux and CCSN rate of Sample A, B, C as shown in Table \ref{tab:local galaxies}. Since the FUV flux of Sample A is arichived from our estimation, we mark it in a dashed red box.} \label{fig:FluxSNe_mixMetallicity}
\end{figure*}
\begin{table}
\caption{Relative ratio of type Ibc SNe to type II SNe ($ {\rm R_{Ibc}/R_{II} }$) under constant SFH.}
\begin{center}
                      \begin{tabular}{l|c|c|c}
                        \hline
                        \hline
                        *& \multicolumn{3}{|c|}{$ {\rm R_{Ibc}/R_{II} }$}\\
                        \hline
                        Observed ratio & \multicolumn{3}{|c|}{$ 0.31 \pm 0.18$}\\
                        \hline
                        \hline
                        Z & single & binary & Mix single \& binary\\
                        \hline
                        0.001 & 0.002& 0.45 & 0.23\\
                        0.004 & 0.07& 0.63 & 0.35\\
                        0.008 & 0.10& 0.60 & 0.35\\
                        0.020 & 0.20& 0.65 & 0.42\\
                        0.040 & 0.23& 0.72 & 0.48\\
                        \hline
                        Mixed Z & 0.22&0.70& 0.46\\
                        \hline
                        
\end{tabular}
\end{center}
\label{tab:SNe rate ratio}
\end{table}
Our constant SFH model is the simplest case to consider. We show how the $ {\rm H\alpha} $ and FUV fluxes vary relative to the CCSN rate in Figure \ref{fig:FluxSNe_mixMetallicity}. The values expected for the 11HUGS galaxies are shown by the boxes indicating the range of values allowed for the observed fluxes and CCSN rates. We find both single-star and binary-star models perform reasonably well with predictions within a factor of 2 to the observed values. Also the agreement is closer for the FUV flux rather than the $ {\rm H\alpha} $ flux. However the models tend to overestimate the amount of $ {\rm H\alpha} $ and FUV fluxes required to reproduce the observed CCSN rate, showing there is a possible excess of CCSNe as found by several studies \citep[e.g.][]{2005PhRvL..95q1101A,2011PhRvD..83l3008K,2012ApJ...756..111M, 2013ApJ...769..113H}. We also find that the standard single-star population achieves the closest fit to the observed parameters. Our binary-star models produce much greater $ {\rm H\alpha} $ flux than the single-star population which causes tension with the observations. However there are limitations to our models which currently do not yet account for the effects of dust on the $ {\rm H\alpha} $ and FUV emission.\\
\\On the other hand there is another test we can apply as shown in Table \ref{tab:SNe rate ratio}. The ratio of the rate of type Ibc SNe to the rate of type II SNe is $0.31 \pm 0.18$. The large error is due to the small number of events. This ratio is strongly dependent on age, interacting binary fraction and metallicity. The effect of the latter is shown in Table \ref{tab:SNe rate ratio}. The predicted CCSN rate ratios for our single-star and binary-star models are either side of the observed value. The single-star prediction is lower than the observed value but within the range allowed by errors. By contrast, the ratio from binary stars is higher than the observed value. However a mix of single-star and binary-star population also gives a good fit to the observed value.

\subsection{Delayed SFH Model}
 \begin{figure}
\centering
\includegraphics[width=255pt]{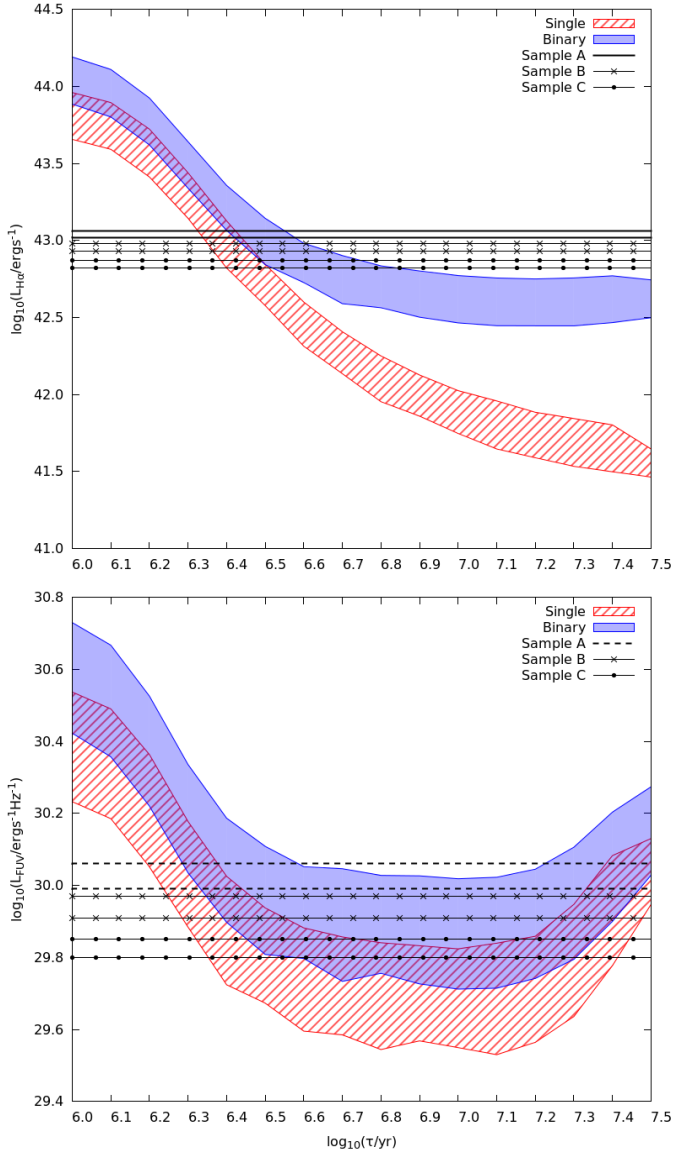}
\centering
\caption{Delayed SFH model: $ {\rm H\alpha} $ and UV flux as a function of timescale parameter $ \tau $. The shaded regions show the range of $ {\rm H\alpha} $ and FUV flux when the predicted CCSN rate matches the observed rate of the 11HUGS galaxies in Table \ref{tab:local galaxies}. The pairs of black straight lines show the observed value range of $ {\rm H\alpha} $ and UV flux in Sample A，B and C as shown in Table \ref{tab:local galaxies}. Because the FUV flux of Sample A is arichived from our estimation, we mark it by two dashed black lines.} \label{fig:Flux_tau_MixZ}
\end{figure}
\begin{figure}
\centering
\includegraphics[width=8.5cm]{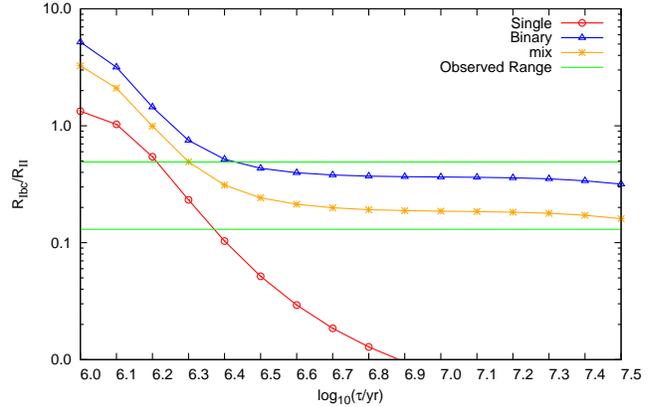}
\centering
\caption{The relative ratio of type Ibc SNe and type II SNe ($ { \rm R_{Ibc}/R_{II}} $) as a function of timescale paparamter $ \tau $. The red dotted line is from sinlge-star model and blue dotted line is from binary-star model. A mixed population of single-star and binary-star model is shown in the yellow dotted line. The green horizontal lines show the observed CCSN ratio range ($ 0.31 \pm 0.18 $) as shown in Table \ref{tab:SNe rate ratio}. } \label{fig:mixZ}
\end{figure}

With constant SFH we predict too much $ {\rm H\alpha} $ and FUV flux relative to the CCSN rate. We therefore decided to investigate alternative SFHs in order to reduce the contribution of young stars. A delayed SFH is applied as described in Section 3. The outcomes we achieve under the delayed star birth assumption are shown in Figure \ref{fig:Flux_tau_MixZ}. Here the shaded regions show the range of $ {\rm H\alpha} $ and FUV fluxes from our synthetic galaxies in which the predicted CCSN rate matches the observed rate of the 11HUGS galaxies.We see that for matching the $ {\rm H\alpha} $ flux and the CCSN rate at the same time the preferred values for $\log(\tau)$ are between 6.3 to 6.7, or times of 2 to 5 Myrs. Although the binary-star models are only just below the observed values and so longer values of $\tau$ would not be unreasonable. While to match the FUV flux and CCSN rate a much wider range of possible values for $\log(\tau)$ are possible between 6.3 to 7.4, or times of 2 to 25 Myrs. This implies that FUV flux follows on a much longer timescale than the $ {\rm H\alpha} $ flux as expected.\\
\\Figure \ref{fig:mixZ} illustrates how the relative ratio of the two CCSN type rates, $ { \rm R_{Ibc}/R_{II}} $, varies with $\tau$. This plot is similar to Figure \ref{fig:Flux_tau_MixZ} showing that as $\tau$ increases the binary-star models are favoured relative to the single-star models where the ratio drops for the single stars. The key difference between the single-star and binary-star models is that the single-star population can only reproduce the observation for $\tau$ in a short period of approximately 2 Myrs. While the binary population, or a mix of single stars and binaries could reproduce the 11HUGS observations over a much longer time scale. However the delayed SFH models are in effect just a starburst model where the age of the burst is linked to its duration. So the best fit we can rule out is that the SFH that best matches the 11HUGS galaxies is one that is not dominated by a recent young burst but older more mature bursts. \\
\\In summary, our delayed SFH models show a preference for 3 Myrs in matching with observed values. This fact could be due to a sample selection effect but this is remote as 11HUGS are a distance limited sample not $ {\rm H\alpha} $ selected. It is more likely that we are over predicting the $ {\rm H\alpha} $ emission over the first 3 Myrs of the star formation.

\section{Discussion \& Conclusions}
We have created a synthetic model of the 11HUGS galaxies
to investigate the effect of interacting
binary stars. We find binaries do not vary the overall CCSN rate to any great degree but do boost the number of type Ibc SNe and the $ {\rm H\alpha} $ and FUV SFRIs linked to hottest stars.\\
\\In trying to match the observations of the 11HUGS galaxies we must discuss the caveats within our population models. We have not yet accounted for the absorption of the ionizing photons by dust grains and therefore these photons do not form $ {\rm H\alpha} $  flux \citep[e.g.][]{2012MNRAS.423.2933R}. This effect is not yet included in our models. In addition, the 11HUGS samples are only nearly complete. As in Sample A there is more star formation occurring than in Sample B, the expected Sample A FUV flux might be greater than we predict here. This result would then be in closer agreement with the binary-star population under constant SFH model. Therefore we should not be too surprised that the match is not perfect but there is more work to refine the synthetic galaxy population.\\
\\The most surprising fact however is that our constant SFH models produce too many CCSNe in the nearest 11 Mpc of space as argued in \citet{2013ApJ...769..113H}. This implies that all supernovae predicted by the models have to be observed. If some massive stars form black holes at core collapse and are unobservable event as suggested by \citet{2014MNRAS.445L..99H} then this excess of stellar deaths compared to stellar population models becomes more severe. However estimating whether a specific stellar model would produce an observable SN or not requires detailed modelling of the core-collapse and the subsequent explosion.\\
\\On the other hand, the delayed SFH model argues that the predicted CCSN rate from $ {\rm H\alpha} $ and FUV SFRIs is dependent on the assumption of the SFH. One interpretation of our preferred delayed SFH model is that it implies we over estimate the flux from young stars and therefore get an excess of CCSNe. A similar result is also provided by ionizing photon leakage and dust attenuation on $ {\rm H\alpha} $ and FUV flux. Therefore to truly determine  if the SFRIs and CCSN rate match, the ionized photon leakage and dust attenuation must be considered. However this is not straightforward. \\
\\We have also investigated how sensitive our results are to the metallicity calibration we use for modelling the metallicities of the 11HUGS galaxies. We find that having the more metal rich biased population leads to small changes in the $ {\rm H\alpha} $ and FUV fluxes. With a more metal poor distribution we would find even more $ {\rm H\alpha} $ and FUV fluxes and would result in greater disagreement between the observations and our models. Therefore the mismatch between our constant star-formation models and observations is difficult to explain by an incorrect metallicity distribution.\\
\\Our main conclusion is that single-star models matches the observations best in terms of the prediction of total CCSN rate, when the effects of ionizing photon leakage and dust attenuation are not considered and assuming the local star formation has followed a constant SFH over the last few 100 Myrs. Binary-star models however prolong the stellar age of stars to go CCSN and therefore provide a better prediction of CCSN rate under various SFH spanning star-burst duration between 3 and 25 Myrs ago. In addition, single-star models under constant SFH predicted a lower ratio of type Ibc relative to type II SN but still within the observed uncertainty. By contrast, binary-star models overestimate the ratio. However, a mix of single-star and binary-star model also gives a good fit as in \citet{{2008MNRAS.384.1109E},{2013MNRAS.436..774E}}. The limitation in making a firm determination is due to the large uncertainty caused by limited numbers of CCSN discovered. With more CCSN discovered within 11 Mpc this uncertainty will decrease. If the number of CCSN is doubled we estimate that it may become possible to rule out the single-star only population. One final argument against the single-star model is that under the delayed SFH the single-star model only match the observations for a burst of star formation only 2 Myrs old. However mixing in binaries to the single-star population provides a better fit with weaker constraints on the age of the population.\\
\\Finally we note that our results are robust if the CCSN sample is complete. However we already obtain too many CCSNe compared to its associated $ {\rm H\alpha} $ and FUV. If there are missing SNe or other massive stellar deaths in our sample that are not included in our CCSN sample then the disagreement between models and observations becomes worse with too many stars dying compared to the number we observe forming.

\section*{ACKNOWLEDGEMENTS}
The authors would like to thank the referee for their helpful comments that have lead to an improved and more robust paper. LX would like to thank the China Scholarship Council for its funding her PhD study at the University of Auckland. JJE thanks the University of Auckland for supporting his research. The authors also wish to acknowledge the contribution of the NeSI high-performance computing facilities and the staff at the Centre for eResearch at the University of Auckland. New Zealand’s national facilities are provided by the New Zealand eScience Infrastructure (NeSI) and funded jointly by NeSI’s collaborator institutions and through the Ministry of Business, Innovation and Employment’s Infrastructure programme. URL: \url{http://www.nesi.org.nz}

\bsp

\label{lastpage}

\end{document}